\documentclass{llncs}
\usepackage{multirow}
 
\usepackage{xspace}

\usepackage{graphicx}
\begin{document}
%
\title{Caring Without Sharing: A Federated Learning Crowdsensing Framework for Diversifying Representation of  Cities}       
\titlerunning{A Federated Learning Crowdsensing Framework}
\author{Michael Cho \and Afra Mashhadi}
\authorrunning{Cho et al.}
 
 \institute{Computing Software System, University of Washington, Bothell,  Washington, USA \\
 \email{mikec87@uw.edu},\email{mashhadi@uw.edu}\\
 }
\maketitle

\begin{abstract}

Mobile Crowdsensing has become main stream paradigm for researchers to collect behavioural data from citizens in large scales. This valuable data can be leveraged to create centralized repositories that can be used to train advanced Artificial Intelligent (AI) models for various  services that benefit society in all aspects.   Although decades of research has explored the viability of Mobile Crowdsensing in terms of incentives and many attempts have been made to reduce the participation barriers, the overshadowing privacy concerns regarding sharing personal data still remain. Recently a new pathway has emerged to enable to shift MCS paradigm  towards a more privacy-preserving  collaborative learning, namely Federated Learning.  In this paper, we posit a first of its kind framework for this emerging paradigm. We demonstrate the functionalities of our framework through a  case study of diversifying two vision algorithms through to learn  the representation of ordinary sidewalk obstacles as part of enhancing visually impaired navigation.  

\end{abstract}

\section{Introduction}



In the past decade the  Mobile Crowdsensing paradigm (MCS) have leveraged power of crowds  for a range of applications to help researchers and  practitioners  enhance their understanding of  the cities and citizens. MCS allows users to participate in a campaign   by providing passive or active data through their sensor enabled mobile devices.  For instance, within the context of smart cities, MCS has enabled the optimization for various \emph{passive sensing} applications, such as pollution, public transportation, traffic congestion, road conditions, etc. Active sensing applications have also helped to advance understanding of the cities through the active contribution of the crowds~\cite{zhongsurvey}. Such applications include FixMyStreet\footnote{https://www.fixmystreet.com} where citizens actively report faults within their neighborhoods or others such as  GeoNotify~\cite{kim2020crowd} where citizen report the sidewalk obstacles that could impact visually impaired navigation.  

To this end, frameworks such as  AWARE~\cite{ferreira2018aware}, Sensus~\cite{xiong2016sensus}, and SensingKit\cite{katevas2016sensingkit} have brought the feasibility of  creating MCS tasks/campaigns to researchers and policy makers. For example AWARE framework allows users to design experiments and run data collection  campaigns that tap into the smartphone sensors  with a few lines of code. Indeed, common to all these frameworks is  that they provide an easy interactive design for the scientific community to design the experiment through a web dashboard, and furthermore they offer storage and communication between the devices and the server.
  
However, there still remains privacy   challenges that could act as participation barriers for  MCS users~\cite{gustarini2016anonymous}.   These privacy concerns are in twofold:    First, user's data could  contain sensitive personal information. Secondly, personal information can be concluded by analyzing the data provided by the user and through   continuous  monitoring. For example, by collecting sensory data related to the user location on the device, the user’s home location information can be obtained. 
 
To overcome privacy concerns,  the crowd-sensing community has recently started to explore alternatives and  possibilities of a paradigm shift that would decouple the data collection and analysis from a centralized approach to a  distributed setting.  To this end, Federated Learning (FL) has emerged as a promising candidate for this paradigm shift~\cite{jiang2020federated}. In FL schema each participant's device holds on to their own data, and a FL server orchastrates a collaborative training by sending the shared model to the devices. In this way, the data always remains local to the client device. 
 
 Inspired by this trend, in this paper we present, FLOAT, a \emph{\bf F}ederated \emph{\bf L}earning framework f\emph{\bf o}r \emph{\bf A}ctive crowdsensing \emph{\bf T}asks. In the design of this framework we pay careful attention in the challenges and opportunities that incur in  this paradigm shift. Our framework relies on Flower~\cite{beutel2020flower} to facilitate FL training on the device end and the orchestration on the server side.  Our framework consists of an interactive dashboard allowing the researchers to setup their task and specify the training properties. On the device end, we design a range of functionalities to enable users to participate in a campaign.  To demonstrate an application of our framework, we present algorithmic and system performance of an obstacle detection use case, where we train state-of-the-art vision models to enhance their representation of ordinary side walk objects.  In summary, we make the following contributions:
  \begin{itemize}
      \item We propose, the first of its kind, an end-to-end framework for bringing FL into crowdsensing tasks.  Our entire framework would be open-sourced and available under Apache 2.0  license for the crowdsensing community to use in their research. 
    
      \item Using FLOAT  as the underlying framework, we present experiments that explore both algorithmic and system-level aspects of federated mobile crowdsensing for an application of Obstacle Detection. We address important research questions as to: {\em How many users and how much data per user is   required to learn representation of 5 ordinary sidewalk obstacles. } 
      
      \item We propose a roadmap that we believe would empower the   research community to address challenges that remain if we are to integrate FL into the MCS paradigm. 

  \end{itemize}


\section{Background and Related Work}

\begin{figure*}
    \centering
    \includegraphics[scale=0.35]{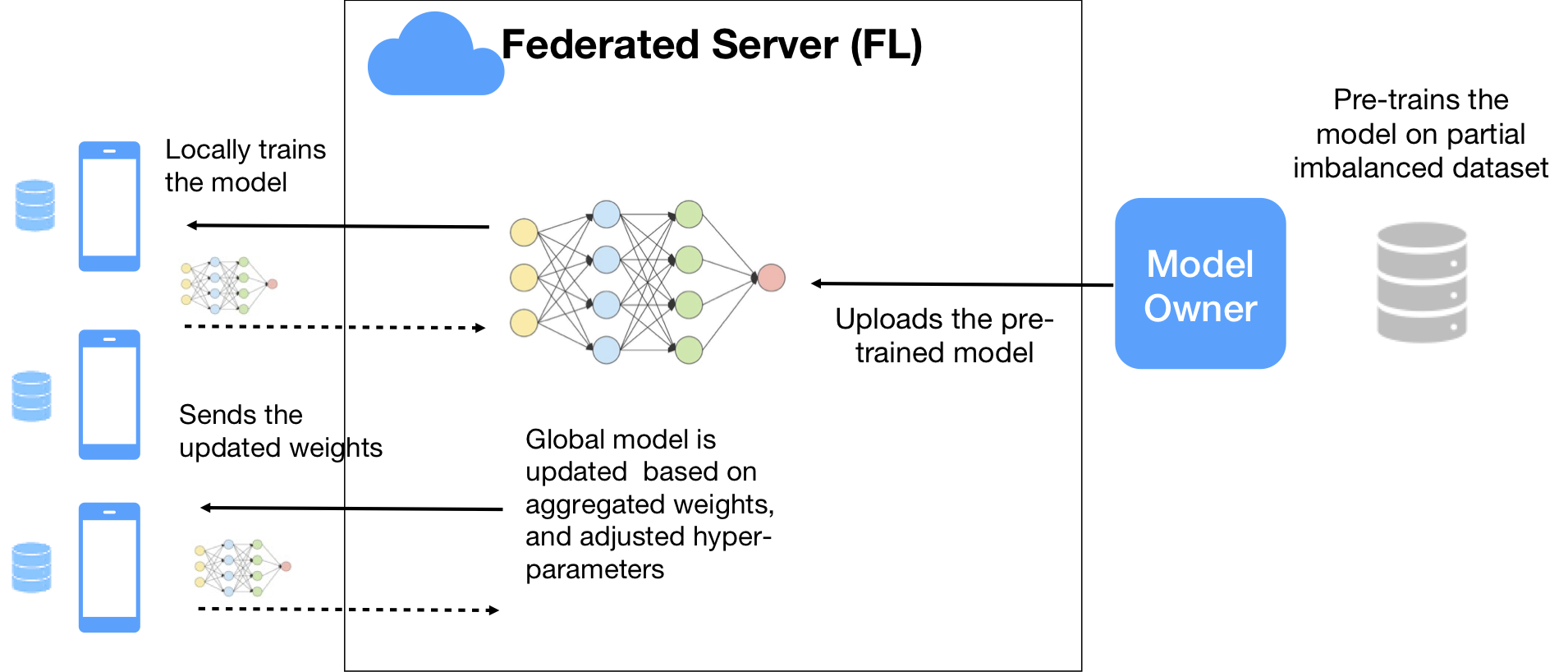}
    \caption{The overview of the interactions between the model owner and the devices under orchestration of the FL server.}
    \label{fig:arch2}
\end{figure*}

 Federated Learning (FL)~\cite{mcmahan2017communication,aledhari2020federated,kairouz2021advances,konevcny2016federated} has been proposed to  provide a privacy-preserving mechanism to leverage de-centralized user data and computation resources to train machine learning models.   Federated learning allows users to collaboratively train a shared model  under the orchestration of a central server while keeping personal data on their devices. There are, in general, two steps in the FL training process (i) local model training on end devices and (ii) global aggregation of updated parameters in the FL server. The training  process of such a FL system usually contains the following three steps as illustrated in Figure~\ref{fig:arch2}:

\begin{enumerate}
\itemsep0em 
    \item The server initializes the global weights, specifies the global model hyper parameters and the training process, and sends the  task to selected participants.
    \item   Participants locally compute training gradients and send the gradients or updated weights   to the server.
    \item  The server performs  aggregation  and shares the new weights with participants.  Steps 2--3 are repeated until the global loss function converges  or a desirable training accuracy is achieved.
\end{enumerate}

To this end, incorporating Federated Learning paradigm into Mobile Crowdsensing tasks can address some of the long existing challenges of MCS and create new opportunities~\cite{jiang2020federated}. In Mobile Crowdsensing the use of personal smart devices that have enough processing capabilities are a prime candidate to integrate with Federated Learning. The benefit of such methodology can be seen in two major aspects. First, Federated Learning helps to preserve the privacy of the user by never uploading the raw collected data. Secondly, it can be leveraged as a means to diversify the representation of the data and lead to more inclusive machine learning models. 

It is worth noting that there also    exists other  great opportunities that arise from this paradigm shift and particularly  from removing the burden of data collection and centralized training.  For instance, in traditional MCS schemes the server must overlook the transmission, and the storage of data. The data is then  pre-processed and used for a centralized training. These processes incur large overhead costs and maintenance and thus reducing those requirements by removing the need for centralized data repositories  would lower the threshold for deployment campaigns. Moreover, MCS incurs large communication costs with the transfer of raw data into the server. This places a burden on the server itself to process this data and takes a large amount of bandwidth from the users as well. By bringing the model to the user and enabling local training, the need for resources in terms of bandwidth is greatly reduced. Finally, in terms of energy consumption, previous work has shown that federated learning can play a large role in reducing CO2 emission associated with training large AI models, by moving the training process into hand-held devices that do not require cooling~\cite{qiu2020can,savazzi2021framework}. Furthermore, previous work has also shown that the energy and memory consumption that are required for on-device training for smart city applications is almost negligible~\cite{beutel2020flower,IJCNN}.

\subsection{Applications of Federated Learning in MCS }
   To date, research in real-world applications of  federated learning are still in  infancy and limited to a handful examples. Smart security~\cite{baig2017future} is an emerging field that has seen most integration of  federated learning in the context of smart cities.   Based on machine learning, smart security can perform post event analysis and self-learning, constantly accumulating experience, and continuously improve pre-warning capabilities. Federated Learning offers a machine learning training scheme that allows the use of the large amounts of collected data in daily applications~\cite{preuveneers2018chained}. Outside of security applications Mashhadi et al. ~\cite{IJCNN} proposed an application of federated learning for discovering urban communities. They showed that by using the GPS traces that are stored on each device, and collaboratively training a deep embedded clustering model, it is possible to detect meaningful urban communities without the need for location information to be shared.



\section{Design Considerations}

In this section  we take a critical view of some of the design challenges that incur due to the suggested paradigm shift and decoupling the data from the centralized approaches.  We address how we design for responding to these challenges in the proposed framework.


\subsection{Challenges}
In order to shift the MCS schema to a decentralized approach where the   participants' data is only held on local devices, various challenges need to be addressed. 
 
\begin{itemize}
    \item {\bf Challenge 1:} Perhaps most challenging aspect of a FL MCS  proposed schema is that it reduces the {\em exploratory} scopes of MCS tasks. Indeed one of the benefits of MCS data collection, has always been on enabling researchers to collect a large volume of data first and then ask what type of research hypothesis could be addressed and which portion of the data is indeed needed to answer those question. In contrast, a federated schema, reduces the scope of this exploratory analysis and enforces researchers to  not only have a very well defined   research question and hypothesis, but most importantly a  {\em model} prior to deployment. 
    
    This property  means that many   phases of exploratory analysis needs to be shifted into a pre-training stage where the model is implemented and trained on a proxy dataset. Such proxy dataset   may or may not be an accurate representation of the participant's data. Indeed, in this vein~\cite{IJCNN} showed that it is possible to {\em pre-}train an urban  community detection model on an aggregated mobility dataset and then dispatch it to be re-trained under the FL setting on client's fine grain location data. 
    
   {\bf Design Goal 1:}   In our framework we design for this functionality by enabling  any pre-trained model and weights to be loaded to FLOAT. Figure~\ref{fig:arch2} presents the proposed schema where the user (i.e., model owner) shares a pre-trained model with the framework which then gets pushed towards the client devices. Furthermore, we also cater for the applications of transfer learning. Transfer learning focuses on storing knowledge gained while solving one problem and applying it to a different but related problem. In many cases where the algorithms are pre-trained to learn a representation of a proxy dataset, it is possible to only   retrain the final layer of the model to account for the specific task. Such approach also significantly reduces the   convergence time and thus reduces the training at the local devices. In FLOAT, we provide   an option to indicate whether the weights of the loaded model are to be fully re-trained or only fine tuned at the  final layer.

\item{\bf Challenge 2:}  A second challenge  in designing for a federated  MCS  schema is the lack of {\em transparency}. That is because participants' data is unseen, it is difficult to assess the outcome of the training. Therefore, questions arise on ``how much contribution did each participant make?'' or   ``How many rounds of training would be needed for the model to converge?''.   

{\bf Design Goal 2:} To address this  challenge,   we design our framework with an interactive interface which  enables the model owner to quantify the outcome of their model  after each round. More specifically we provide two ways of validation:

{\bf Server Validation}: The model owner is able to specify a path to to a centralized dataset that could be used to evaluate the accuracy of the updated model after each round of training. 

{\bf Client Validation}: The validation is entirely on the clients devices. For the cases that the model owner does not have a centralized validation dataset to validate the updated model against it, our framework enables client validation where a portion of participants are selected for validating the model. 

\end{itemize}

 \subsection{Opportunities}
 
There  exists multiple  great opportunities that arise from removing the burden of data collection and centralized training.  For instance, in traditional MCS schemes the server must overlook the transmission and storage of data. The data is then  pre-processed and used for a centralized training. These processes incur large overhead costs and maintenance. Reducing those requirements by removing the need for centralized data repositories  would lower the threshold for deployment campaigns. 

Moreover, MCS incurs large communication costs with the transfer of raw data into the server. This places a burden on the server itself to process this data and takes a large amount of bandwidth from the users as well. By bringing the model to the user and enabling local training, the need for resources in terms of bandwidth is greatly reduced.

Finally, in terms of energy consumption, previous work has shown that federated learning can play a large role in reducing CO2 emission associated with training large AI models, by moving the training process into hand-held devices that do not require cooling~\cite{qiu2020can,savazzi2021framework}. Furthermore, previous work has also shown that the energy and memory consumption that are required for on-device training for smart city applications is almost negligible~\cite{IJCNN}.  


\begin{figure*}
\centering
 \includegraphics[scale=0.45]{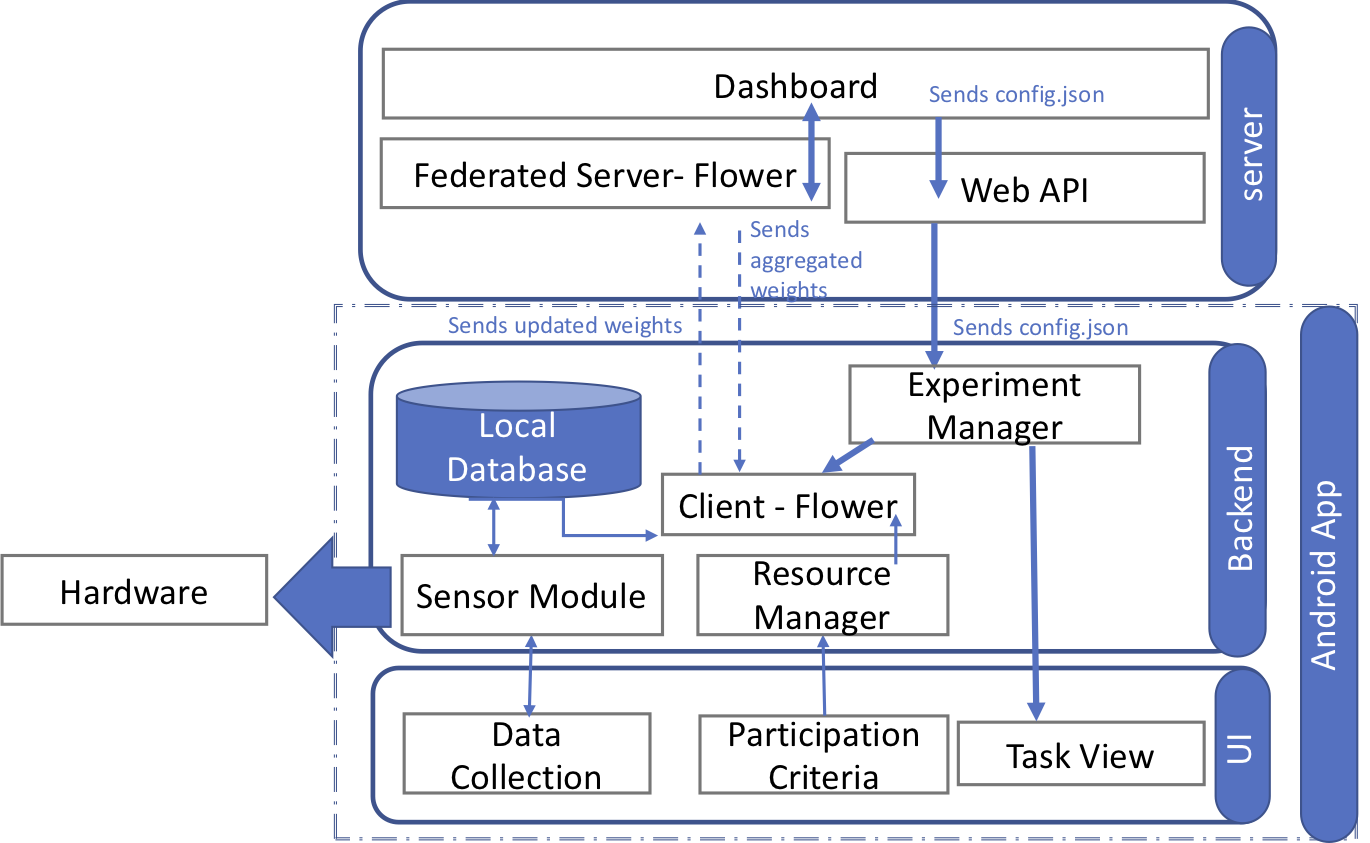}
    \caption{Architecture of FLOAT depicting the server and the client part of the  framework.}
    \label{fig:arch}
\end{figure*}

\section{Overview of FLOAT Framework}

Figure~\ref{fig:arch} presents the overall architecture design of our framework. As can be seen our framework consist of the server and the client end. Both these parts of the framework rely on Flower~\cite{beutel2020flower} as an underlying Federated Learning platform. We thus first describe Flower for the sake of clarity before moving on to  describe each component of our framework and the interactions amongst them.

\subsubsection{Flower.}

Our framework relies on Flower~\cite{beutel2020flower} as an agnostic and scale-able solution for federated learning.  Flower offers a stable, language and ML framework-agnostic implementation of the core components of a federated learning system. In particular by using Flower as an underlying FL   implementation, FLOAT is able to inherit the agnostic properties of Flower. That is our framework is  able to support Machine Learning Models written in either Tensorflow or Pytorch.  

Furthermore, Flower allows for rapid transition of existing ML training pipelines into a FL setup to evaluate their convergence properties and training time in a federated setting. This includes various strategies from aggregation methods (e.g., FedAvg~\cite{mcmahan2017communication}, QffedAvg~\cite{li2019fair}, and many others) and evaluation methods.  

Most importantly, we chose Flower as it provides support for extending FL implementations to mobile and wireless clients, with heterogeneous compute, memory, and network resources  (e.g., phone, tablet, embedded) and thus ideal for the smart city applications.

\subsection{Server}

 The back-end server of our framework is responsible for taking the design of the experiment from the end user, communicating it with devices, and initiating the experiment. 

\begin{figure}[!ht]
\centering
 \includegraphics[scale=0.35]{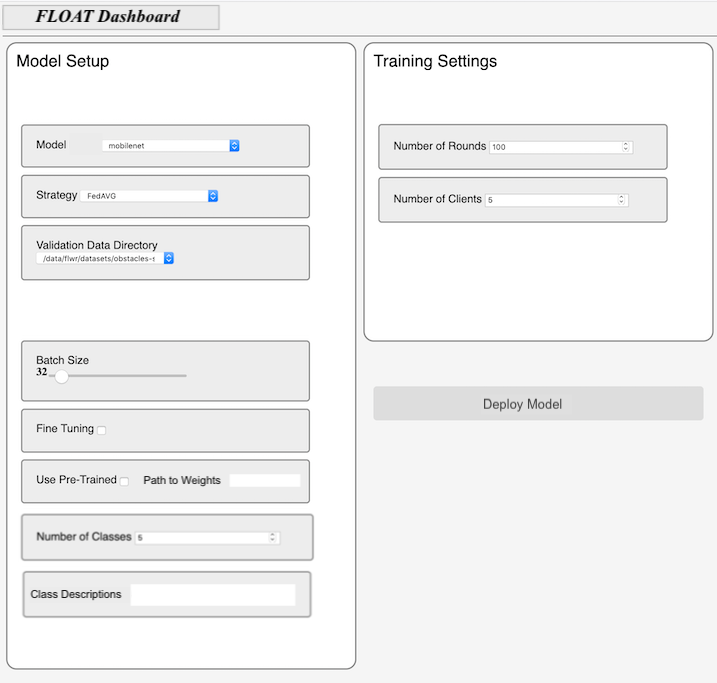}
    \caption{Part of the FLOAT Dashboard that acts as an interface with the user to set up the task.}
    \label{fig:dashboard}
\end{figure}

{\bf Dashboard.} To facilitate an easy interaction our framework has an   interactive dashboard (Figure~\ref{fig:dashboard}) that enables researchers to load their own model and specify the training parameters such as the number of training devices (Design Goal 1), the number of data points per each participant, and the hyper-parameters of the experiment. We developed this dashabord in python-based Flask server~\cite{flask} which enables  us to build up the  web-application and easily modify the components of the interface. 

\begin{figure*}[!ht]
\centering
 \includegraphics[scale=0.3]{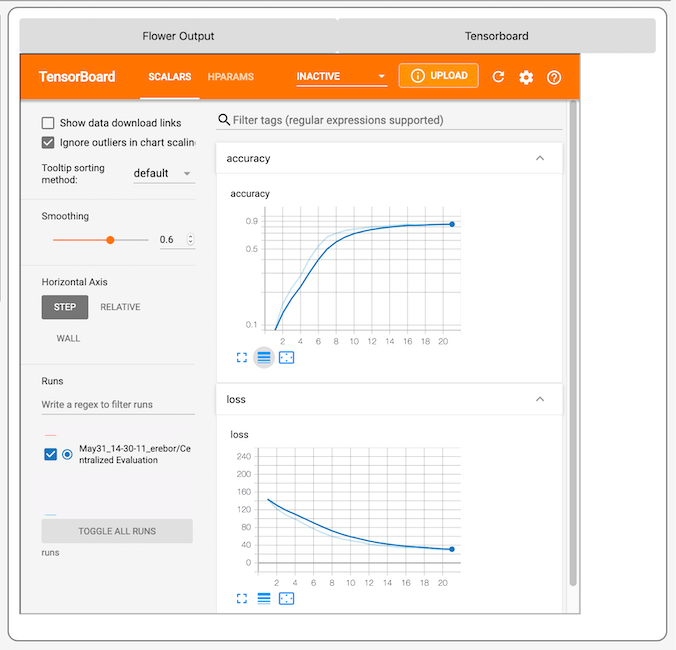}
    \caption{The output component of the dashboard visualizing live results of the training.}
    \label{fig:tb}
\end{figure*}

In addition to taking input from the user, FLOAT dashboard is designed to provide transparency into training (Design Goal 2) by integrating a live visualization dashboard implemented in TensorBoard~\cite{tb}. Figure~\ref{fig:tb} presents this component of the dashboard. An alternative tab on the dashboard allows the users to switch to a debugging interface where the underlying Flower messages presenting INFO, DEBUG, and ERROR that are happening during the training round as observed by the RPC communication channel are displayed.  

{\bf FL server.} To start an instance of the FL server, the information from the dashboard is directly communicated to  the FL server python code.  More specifically these are the following parameters: i) rounds of training; ii) aggregation strategy, iii) evaluation strategy and path to a validation dataset (if applicable), and iv) number of clients. 
The FL (Flower) server  configures the next round of FL by sending the required configurations to the  clients via  bi-directional gRPC channel, receives the resulting client updates (or failures) from the clients using the same gRPC, and   aggregates the results using  the strategy.

{\bf Web API module.} Additionally in order to  communicate the experiment setting with the clients devices, the settings that are entered in the dashboard are saved   as a {\em config.json} and sent directly to the clients device.  This configuration file includes information regarding the hyper-parameters of the model and local training instructions: i) the model and the initialized weights (if applicable), ii) instructions on whether the model is to be fully retrained or fine-tuned, iii) number of data points per class, iv) finally number and description of classes.

\subsection{Client}

The client component of the FLOAT is currently designed to support Andriod devices. As depicted in Figure~\ref{fig:arch}, this component has two main parts: the Backend and the User Interface. We describe the interaction between each part next. 

\subsubsection{Backend Modules}
The modules in the backend part of the app are responsible to bridge between the FL server and the user interface. These are {\em Experiment Manager}, {\em FL Client}, {\em Resource Manager} and finally the {\em Sensor Module}.

 {\bf Experiment Manager.} As depicted in Figure~\ref{fig:arch}, this  module is in charge of receiving the training instructions, parameters, and model from the server. It then communicates the information about the task to the participant through the Task View component of the UI. It is also responsible for setting up the   FL client  and initializing the training.   

{\bf FL Client. } The FL client is responsible for training the shared model. As described earlier we use Flower as the underlying framework to support the on-device training. The flexibility of the flower architecture enables our framework to receive any models and weights and simply assemble a FL client code. Furthermore because Flower is designed for heterogeneous devices, the FL client module can be later integrated into different devices and platforms (e.g., iOS).    As depicted in the Figure~\ref{fig:arch}, the FL client code also has direct access to user's local dataset. FL Client connects to the gRPC channel  which is responsible for monitoring these connections and for sending and receiving Flower Protocol messages.

{\bf Resource Manager.} The resource manager is responsible for monitoring the current device resources and communicating that with the flower client and server.


{\bf Sensor Module.}  This module is responsible for  enable specific data to be used as part of the training tasks. This is done by direct communication with the Data Collection Module of the UI as we describe next.

\subsubsection{User Interface}
~\\
Through the designed UI,   participants are able to view the information and description of the MCS task and provide consent in taking part in the campaign.  Furthermore, we design the UI with the vision of enabling the participant to indicate their participation criteria such as minimum available resources or stable connectivity (WiFi). Finally the Data Collection module of the UI enables the participant to specifically collect data that is required for the active MCS task. This module therefore directly communicates with the sensor module of the client backend and has access both to hardware resources (e.g., camera, GPS etc) and the on-device database.  

\section{Case Study: Obstacle Detection}

In this section we present one of many possible use cases for our framework. The use case here is motivated by GeoNotify an application that was designed to assist visually impaired with navigation~\cite{kim2020crowd}. GeoNotify is a  crowd-sourcing application that notifies the users about the obstacles on the sidewalks and re-routes them to avoid those obstacles using audio messages. To map the obstacles the system relies on crowdsourced information from the users by enabling them to take a photo of sidewalk obstacles and an audio description and submitting it to a centralized repository.  The underlying back-end server learns: 1) the GPS location of the obstacle and uses this information to reroute other users; 2) the image representation of the obstacle. The focus of this section is on the latter component of the system which aims to retrain vision models to learn   representation of everyday sidewalk obstacles.

\subsection{Obstacle Detection Model}

A large scale study by the Royal Institute of Blind People~\cite{ribp} interviewed  500 visually impaired participants   for over three months. As part of this report the institute highlights that often the ordinary sidewalk obstacles are the most common causes of injury daily navigation of those with partial vision impairment. This report identifies  five most common obstacles that impact visually impaired daily: cars parked on the pavement,  advertising-boards, bins and recycling boxes on pavements, street furniture (such as chair and tables). Some of these classes (e.g., chair and table) are common objects that are labeled in   image recognition datasets such as ImageNet~\cite{deng2009imagenet} and thus are detectable using existing Convolutional Neural Network (CNN) models.  However,  the presentations of these common objects across the world can vary significantly. Others such as boards, sidewalk signs and potholes  are specific to the context of this report and are not present as part of ImageNet. Indeed, earlier studies showed that the accuracy of the   state-of-the-art models in detecting common sidewalk obstacles ranges between 10-40\%~\cite{kim2020crowd}. 

To enable vision models to learn a diverse representations of the sidewalk objects, we experiment with two state-of-the-art light weight models, namely  MobileNet~\cite{howard2017mobilenets} and SqueezeNet~\cite{iandola2016squeezenet}. Furthermore, as the system is designed to rely on the crowds' contribution  to enhance these models,   it is important to quantify the number of images that are required to train the models to learn the representation of the sidewalk objects.  We thus seek to   answer the following research questions:

\begin{itemize}
    \item RQ1: How many participants are needed to collaboratively train each model? 
    \item RQ2: How many photos per participant is required to enhance the model?
    \item RQ3: How much resources per client device is required to collaboratively train the model?

\end{itemize}

\subsection{Experiment Setting}

In order to answer the above research questions, we use FLOAT as underlying framework to simulate the described active crowdsensing task amongst devices. The details of our experiments are as follows:

{\bf Data.} We used the manually curated dataset as published by~\cite{kim2020crowd}. We segmented the training portion of this dataset into equal and uniform distribution amongst the clients in an IID fashion. That is each client has an equal number of photos per class. Our dataset consists of 5 classes  with a varying  number of   images, $x$, per class. We kept the validation part of this dataset, which includes 100 images per each class, as the validation dataset on the server.

{\bf Training.}  We experiment with    pre-trained MobileNet and SqueezeNet models (weights of pretraining on ImageNet~\cite{deng2009imagenet}) and study the impact of transfer learning of each model on algorithmic and system performance. To do so we freeze the learning on the earlier layers of the model where the weights are transferred from the pre-trained weights and then retrain the final layer of the network (i.e., the fully connected layer) to learn the representations of our domain specific images. We trained the each model for 20 rounds under the FL setting. During each of the 20 rounds, clients perform one epoch of SGD (batch size 32, momentum 0.9 and learning rate of 0.001) before sending the updated model parameters back to the server. The aggregation strategy was configured as FedAvg. 
 

 {\bf Platform.} We ran an instance of the FLOAT framework on  a local server with 4    GPU GeForce RTX 2080 Ti with CUDA 9.   For answering the first two research questions we \emph{simulated} the clients on the same server each with their own data private repositories. To answer the third research question and compute the system performance on the mobile devices, we ran an end-to-end instance of training on an Android device and measured the system usage of this device. 
 


\begin{figure*}[!ht]
\centering
    \includegraphics[scale=0.2]{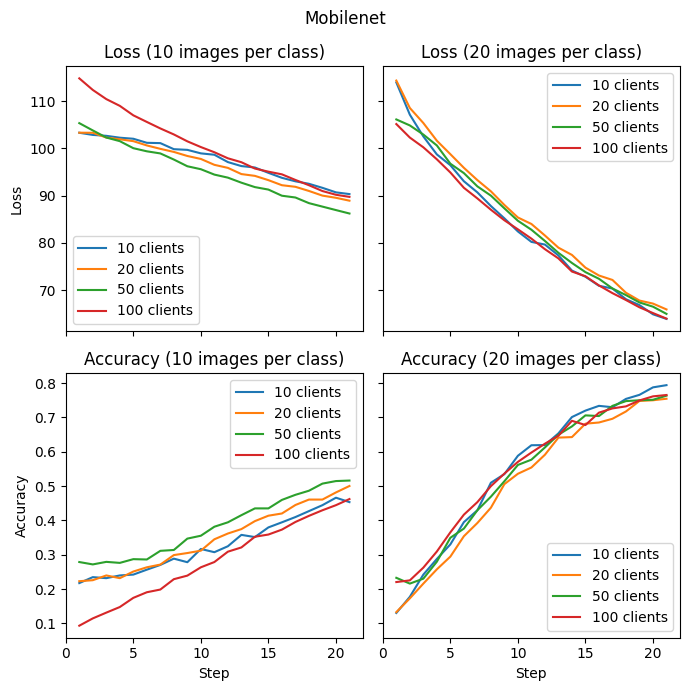}
    \includegraphics[scale=0.2]{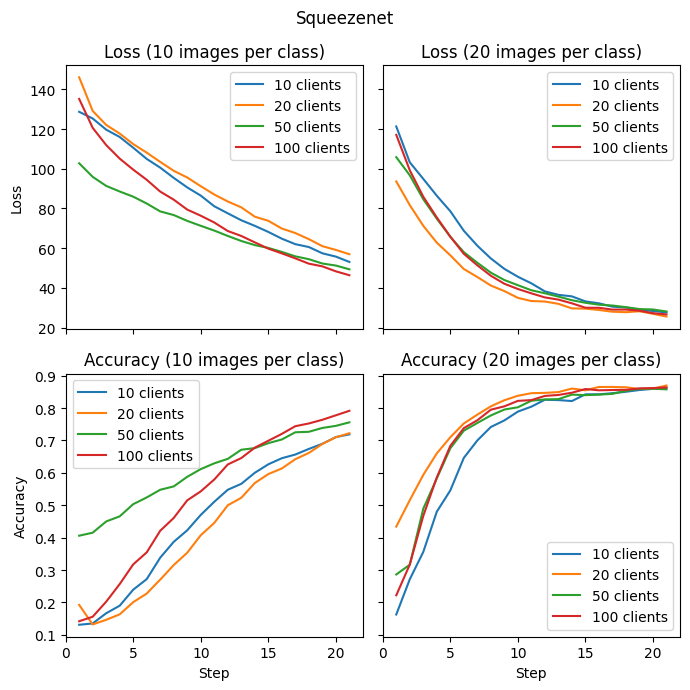}
    \caption{Results of the algorithmic performance of MobileNet (left) and Squeezenet(right) for varying number of clients and images.}
    \label{fig:res}
\end{figure*}

\subsection{Experimental Results}

In this section we present the algorithmic performance and system measurements of the described study.

\subsubsection{Algorithmic Performance}

Figure~\ref{fig:res} reports on the algorithmic performance of MobileNet and Squeezenet respectively for varying number of clients and varying data size.  To answer the first research question on how many participants are needed to collaboratively contribute to the model, we can see that as little as 10 clients can improve the accuracy of the model to 0.4-0.8 for MobileNet and 0.6-0.8 for SqueezeNet.  To answer the second research question on how many images ($x$) each participant requires to have, we can observe that  the more images that users have (per class) the quicker the models converges. Particularly in the case of SqueezeNet we can see that the accuracy {\em exponentially} increases for $x=20$, making the number of the participants not as important of a factor as when $x=10$.  For MobileNet we see that the model performes very poorly in learning the representation of the objects when participants images is $x=10$ and after 20 rounds of training it leads to only 0.5 accuracy. SqueezeNet on the other hand reaches 0.8 accuracy with the larger number of clients ($n=100$).

\subsubsection{System Performance}

\begin{figure*}[!ht]
    \centering
    \includegraphics[scale=0.27]{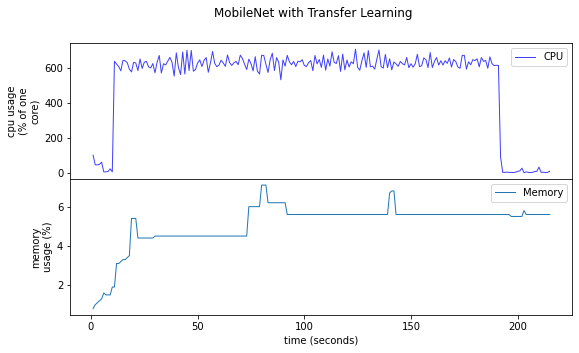}
    \includegraphics[scale=0.27]{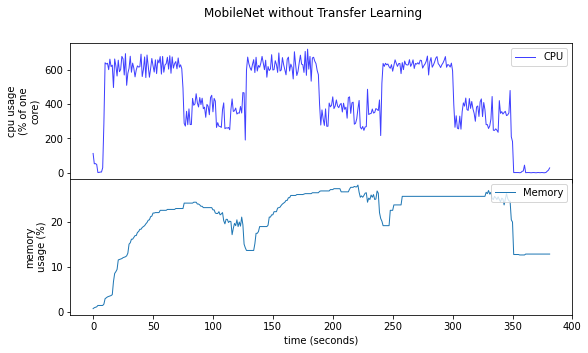}
    \caption{CPU and Memory usage of one round of training of MobileNet on Samsung S21 Ultra with and without Transfer Learning.}
    \label{fig:system}
\end{figure*}

Finally in terms of resources we measured memory, CPU, and energy consumption of the described experiments on a Samsung S21 Ultra, with 8GB RAM and  8 CPU cores.  We experimented   running each model  with and without Transfer Learning (TF)  on this device using a training dataset that is composed of 20 pictures per class. The client only performed the training and no validation was done on the client. Figure~\ref{fig:system} illustrates the CPU and memory usage for MobileNet with and without TF respectively. 
We observe that the maximum memory usage during the entire training remains low. Indeed our comparison of the two models suggest that the memory usage is approximately between 300MB-700MB (3-7\% of the total RAM) for Squeezenet and MobileNet respectively when the models are leveraging transfer learning.  Table~\ref{tab:system} shows the memory and training time for both models. In terms of energy consumption, one round of training (of MobileNet without TL) consumed 10.3 mAh on average that corresponds to less than 0.15\% of the total available battery on a fully charged device. 
 
 \begin{table}[!ht]
    \centering
    \begin{tabular}{ |c|c|c| }
    \hline
         Model &   Time (per one round) & Max Memory   \\
         \hline
         \hline
         MobileNet (TL) &   180 s & 728 MB \\
         \hline
         MobileNet (No TL) &   340 s & 2.8 GB\\
         \hline
         SqueezeNet (TL) &  7 s & 384 MB\\
         \hline
         SqueezeNet (No TL) &  32 s  & 1.2GB\\      
         \hline
    \end{tabular}
    \caption{On device measurements of memory usage and training time for one round of local training. }
    \label{tab:system}
\end{table}
 
 Bringing these results together with the algorithmic performance we can see that   our framework is a viable option for the described MCS task and can enhance the   representation of the objects in  as few as 10 rounds of training with very little participation burden (number of images and computational burden of the devices (time, energy, memory, and CPU).

\section{Discussion}

In this section we first describe some limitation and future extension to our work before putting forward  a future road map on what we believe are the important challenges for the research community to tackle to ensure the adoption of the federated learning beyond theory and opportunities for integration in real-world studies.

\subsection{Limitation}

One  limitation  of the  evaluation presented here is that the participation of  users was simulated and we did not evaluate our framework on  images {\em collected by} actual users who may present a more diverse representation of the objects across the world. Furthermore, as these images were collected from Internet, they might have a have higher quality and fail to represent the  heterogeneity of the smart-phones devices, and the impact it might impose on the training.  

Finally, although we designed our framework with paying particular attention to some of the design challenges, there are others that we have not addressed yet. One of such challenges is the impact of the {\em noisy labels}. That is where the participant intentionally (i.e., poison attacks) or unintentionally associate a wrong label with the input data. Noisy label detection is a hard task as they are ubiquitous in the real world and prominent in   crowd-sourcing applications~\cite{yan2014learning,yu2018learning}.  Deep neural networks, including CNNs, have a high capacity to fit noisy labels~\cite{zhang2016understanding}. That is,  these models  can memorize \emph{easy} instances first and gradually adapt to hard instances as training epochs become large. When noisy labels frequently exist, deep learning models will eventually memorize these wrong labels, which leads to poor generalization performance. In our future work we plan to tackle this challenge by integrating a $Trust Module$ in both the client side (in order to  detect noisy labels) and on the  server side (in order to select reputable participants).  



\subsection{Research Road Map}

{\bf User Acceptance:}  We believe first and foremost important challenge with the proposed paradigm shift is to study and assess users' acceptance. Great amount of literature exists on user's privacy concerns in various platforms from Social Media~\cite{smith2012big,beigi2020survey} to IoT~\cite{badii2020smart,liu2018survey} and  MCS~\cite{gustarini2016anonymous,diamantopoulou2018assessment}. However, little is known on how users will perceive systems that rely on FL schema. Would such schema actually ease their privacy concerns? What type of awareness and education is needed for the users to understand the underlying benefits of such schema?   We believe this is an urgent qualitative research question that the research community needs to answer if we are to see more real-world use cases and applications of federated MCS.

{\bf Participants Selection: } The vast majority of federated methods  are passive in the sense that they do not aim to influence which devices participate, or only select the participants based on the available resources (e.g., connected to WiFi and battery level). We believe  that if we are to use a federated MCS in real-world use cases, it is important to account for the diversity not only at the point of aggregation but at the point of participant selection.  We believe more research initiative is needed to  enable participant selection based on alternative metrics, such as fairness criteria~\cite{barocas-hardt-narayanan,binns2020apparent}, that could enable increasing fairness of the overall model.

{\bf Human Data Interaction:} Finally we believe important conversations and debates are needed to take  place if FL systems are to become more applicable in training ML models. For instance how to adapt policies such as   {\em right to be forgotten} to federated schema. In other words, ``how can a model forget  users' contribution to it''. We believe  stepping away from centralized datasets, brings all new set of regulatory challenges and now is the crucial time for the research community  to start on creating forums for these types of conversations.  


\section{Conclusion}

In this paper we presented FLOAT, a federated learning framework for adapting active MCS tasks into a privacy-preserving FL schema. Through a use case, we demonstrated how FLOAT can be used to  learn a diverse representation of  sidewalk obstacles  using as little as 10 images per object and 20 rounds of training. We also demonstrated the viability of our proposed framework through measuring resource consumption on an ordinary smart phone.

\bibliographystyle{splncs04.bst}
\bibliography{ref.bib}

\end{document}